\documentclass[conference]{IEEEtran}
\IEEEoverridecommandlockouts
\usepackage{bm}
\usepackage{cite}
\usepackage{amsmath,amssymb,amsfonts}
\usepackage{mathtools}
\usepackage{algorithmic}
\usepackage{graphicx}
\usepackage{textcomp}
\usepackage{xcolor}
\usepackage{subfigure}
\usepackage{hyperref}
\DeclareMathOperator*{\argmax}{arg\,max}

\def\BibTeX{\rm B\kern-.05em{\sc i\kern-.025em b}\kern-.08emT\kern-.1667em\lower.7ex\hbox{E}\kern-.125emX}

\newcounter{MYtempeqncnt}

\begin{document}
\title{Beam Tracking for Dynamic mmWave Channels: A New Training Beam Sequence Design Approach}
\author{\IEEEauthorblockN{Deyou Zhang, Ming Xiao, and Mikael Skoglund}
\IEEEauthorblockA{\textit{School of Electrical Engineering and Computer Science, KTH Royal Institute of Technology, 10044 Stockholm, Sweden} \\ email: \{deyou, mingx, skoglund\}@kth.se}}

\IEEEspecialpapernotice{(Invited Paper)}

\maketitle

\begin{abstract}
In this paper, we develop an efficient training beam sequence design approach for millimeter wave MISO tracking systems. We impose a discrete state Markov process assumption on the evolution of the angle of departure and introduce the maximum a posteriori criterion to track it in each beam training period. Since it is infeasible to derive an explicit expression for the resultant tracking error probability, we turn to its upper bound, which possesses a closed-form expression and is therefore leveraged as the objective function to optimize the training beam sequence. Considering the complicated objective function and the unit modulus constraints imposed by analog phase shifters, we resort to the particle swarm algorithm to solve the formulated optimization problem. Numerical results validate the superiority of the proposed training beam sequence design approach.
\end{abstract}


\section{Introduction}
Millimeter wave (mmWave) communications have received great attention from both academia and industry thanks to the huge available spectrum resources at mmWave frequency bands and the resultant multi-gigabit data transmission rate \cite{Ming-Survey}. However, to realize such an appealing performance, directional beamforming with a large antenna array is essential \cite{RH-Survey, Roh-Beamforming}, which requires accurate channel state information (CSI). Thus, mmWave systems usually consume a lot of pilot resources to estimate their CSI, incurring significant training overhead. This condition becomes even worse in time-varying scenarios where the time-demanding channel estimation process has to be invoked periodically to update the stale CSI. Consequently, developing an algorithm that can do channel estimation with low overhead is crucial for mmWave systems, especially when their channels are time-varying.

Since a typical mmWave channel consists of only a few dominant paths and each path can be characterized by its angle of departure (AoD), angle of arrival (AoA), and path gain coefficient, it is sufficient to only estimate these channel parameters rather than the large dimensional channel matrix \cite{RH-MSHC}. Furthermore, as one transmission time interval (TTI) in mmWave networks is as short as 0.1 milliseconds \cite{mmWave-URLLC}, the channel parameters between two successive TTIs can be correlated if no abrupt events such as blockages occur, as reported by \cite{Patra-Tracking, Palacios-Tracking}. This guarantees that the AoDs, AoAs, and path gain coefficients estimated in previous TTIs can be leveraged to accelerate the current beam training process\footnote{The AoA and AoD estimation process is often referred to as beam training in mmWave communications.}\cite{Patra-Tracking}. Such prior information aided beam training is also known as beam tracking, which has been popular in the state-of-the-art mmWave researches \cite{BT-Survey}.

In particular, to leverage the sparse nature of mmWave channels \cite{mmWaveChannelModeling}, a lot of angular space based beam tracking algorithms have been reported, e.g., \cite{Duan-Tracking, MAP, POMDP}. As shown in \cite{Duan-Tracking}, by ignoring the grid quantization errors, the problem of estimating AoA-AoD pairs in each beam tracking period can be formulated as a sparse recovery problem. By leveraging an approximate message passing algorithm, the indices of nonzero entries in the sparse virtual channel matrix are estimated, which uniquely reflect the current AoA-AoD pairs. Moreover, to guarantee an appealing beam tracking performance at the low SNR range, several directional training beam sequence design approaches were also investigated in \cite{Duan-Tracking}. To incorporate the channel dynamics into training beam sequence design, the authors in \cite{MAP} defined a weighted Cramer-Rao lower bound, and the training beam sequence design problem in this literature was solved by selecting a set of beam codewords that can minimize their defined metric. Furthermore, the authors in \cite{POMDP} proposed to transform the beam tracking problem into a partially observable Markov decision process (POMDP) problem. Within their POMDP framework, both optimal and suboptimal codebook based training beam sequence design policies were derived. Despite the good performance of those directional training beams in the cases when AoDs change smoothly, their performances will degrade heavily when AoDs change significantly.

To address the aforementioned issue, we develop a non-codebook based training beam sequence design approach for efficient beam tracking. Unlike the highly directional codebook based training beams, the beampatterns of the employed beamformers are more flexible and can be optimized to adapt to different beam tracking scenarios. To be more specific, we consider a dynamic mmWave multiple-input single-output (MISO) channel and impose a discrete Markov process assumption on the evolution of its AoD. To incorporate the AoD transition information into beam tracking, we propose to leverage the maximum a posteriori (MAP) criterion for AoD estimation in each beam tracking period. Because it is infeasible to derive an explicit form of the resultant tracking error probability (TEP), we turn to its upper bound, which possesses a closed-form expression. We rely on this upper bound as the objective function to optimize the training beam sequence employed in each beam tracking period. Considering the complicated objective function and the unit modulus constraints imposed by analog phase shifters, we resort to the particle swarm algorithm (PSA) to solve the formulated optimization problem, earning a set of efficient training beams for AoD tracking\footnote{In this paper, we use the two terms AoD tracking and beam tracking interchangeably.}. Simulation results validate the superiority of the proposed training beam sequence design approach, particularly in the cases when the AoD changes significantly between two consecutive beam tracking periods.

\textbf{Notations}: Bold uppercase letters (e.g., $\bf A$), bold lowercase letters (e.g., $\bf a$), and non-bold letters (e.g., $a$) are respectively used to indicate matrices, vectors, and scalars. $[\mathbf A]_{i, :}$, $[\mathbf A]_{:, j}$, $[\mathbf A]_{i, j}$, $\mathbf A^{\ast}$, $|\mathbf A|$, $\mathbf A^{\rm T}$, $\mathbf A^{\rm H}$, and $rank(\mathbf A)$ denote the $i$-th row, $j$-th column, $(i,j)$-th entry, conjugate, determinant, transpose, conjugate transpose, and rank of $\bf A$, respectively. $\|\cdot\|$ is the $\ell_2$-norm operator. $\mathbf I$ is the identity matrix; $\mathbf e$ is the standard basis vector; and $\mathbf 0$ is the zero vector. $\mathbf x \sim {\cal CN}(\mathbf m, \mathbf M)$ represents a complex Gaussian vector with mean $\bf m$ and covariance matrix $\bf M$. $\Pr\left\{E_1 \mid E_2\right\}$ is the probability of event $E_1$ conditioned on event $E_2$. $\propto$ indicates proportionality up to a constant scale factor. $\Re\{\mathbf a\}$ and $\Im\{\mathbf a\}$ are the real and imaginary parts of $\bf a$, respectively.

\section{System Model}\label{SM}
We consider a mmWave MISO system, where one base station (BS) equipped with $N_{\rm T}$ antennas communicates with a single-antenna user\footnote{Motivated by \cite{Zhao-JSAC}, the proposed beam tracking approach can be directly extended to multi-user scenarios leveraging narrowband frequency tones. To be more specific, by assigning each radio frequency chain a unique frequency tone to individually serve one user, the multi-user beam tracking problem is decoupled into multiple single-user beam tracking problems.}. Referring to \cite{mmWaveChannelModeling}, since only a few scattering clusters exist between mmWave transceivers, the considered MISO channel can be thus modeled as \cite{DJ-TWC}
\begin{equation}\label{MCH-1}
    {\bf h} = \sqrt{N_{\rm T}} \sum\limits_{i = 1}^{L} \alpha_{i} {\bf a}_{\rm T}(\theta_{i}),
\end{equation}
where $L$ denotes the total number of propagation paths and the two variables $\alpha_{i}$, $\theta_{i}$ respectively denote the gain coefficient and the normalized AoD of the $i$-th path, $\forall i \in \{1, \cdots, L\}$. ${\bf a}_{\rm T}(\theta_{i})$ is termed antenna array response vector. By assuming a uniform linear array at the BS, we can express ${\bf a}_{\rm T}(\theta_{i})$ as
\begin{equation*}
    \mathbf a_{\rm T}(\theta_{i}) = \frac{1}{\sqrt{N_{\rm T}}} \left[1, \exp\{j \theta_{i}\}, \cdots, \exp\{j (N_{\rm T} - 1) \theta_{i}\} \right]^{\rm T}.
\end{equation*}
The relationship between $\theta_i$ and the physical AoD, denoted by $\phi_{i}$, $\forall i \in \{1, \cdots, L\}$, is given by
\begin{equation*}
    \theta_{i} \triangleq \frac{2 \pi d \cos(\phi_{i})} {\lambda_f},
\end{equation*}
where $d$ is the spacing between two adjacent antenna elements at the transmitter, and $\lambda_f$ is the carrier wavelength. Moreover, we assume that $\phi_{i} \in [0, \pi]$ and $d = 0.5 \lambda_f$, such that $\theta_{i} = \pi \cos(\phi_i) \in [-\pi, \pi]$, and there exists a one-to-one mapping between $\theta_{i}$ and $\phi_{i}$.

\begin{figure}
    \vskip 2pt
    \centering
    \includegraphics[width = 8.2cm]{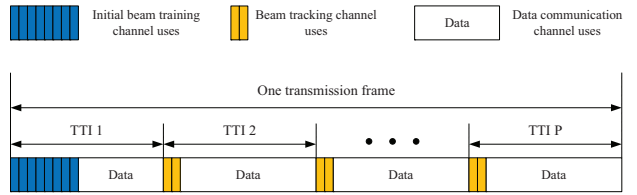}
    \caption{Frame structure of the employed beam training protocol.}\label{BT-Protocol}
\end{figure}

Following \cite{RH-MSHC, POMDP, MAP}, in this paper we assume that $\theta_i$'s are taken from a uniform grid of $N$ points\footnote{As reported in \cite{RH-MSHC}, \cite{POMDP}, and \cite{My-TCOM}, the grid quantization errors only lead to minor performance loss when $N \ge 2N_{\rm T}$.}, given by
\begin{equation}\label{MCH-3}
    \bar \theta_n =  \frac{\pi}{N} + \frac{2 \pi(n - 1)}{N}, ~\forall n = 1, \cdots, N,
\end{equation}
where $N > N_{\rm T}$ to guarantee a fine angular resolution. To reflect the dynamic scenario, we further assume that $\theta_i$ changes within $\{\bar \theta_1, \cdots, \bar \theta_N\} \triangleq \Theta$ following a discrete Markov process. An example of the AoD transition function between two successive TTIs is given by
\begin{eqnarray}\label{AD-1}
    \Pr\left\{\theta_{i,t} = \bar \theta_{\kappa_i} ~|~ \theta_{i, t - 1} = \bar \theta_{\iota_i} \right\} ~~~~~~~~~~~~~~~~~~~ \nonumber \\[2ex]
    = \begin{dcases}
        C_0 \beta^{|\kappa_{i} - \iota_{i}|}, & |\kappa_{i} - \iota_{i}| \le \sigma, \\[2ex]
            ~~~~0, & |\kappa_{i} - \iota_{i}| > \sigma, \\
    \end{dcases}
\end{eqnarray}
where $\iota_i, \kappa_i \in \{1, \cdots, N\}$, and the parameter $\beta \in [0, 1]$ characterizes the variation extent of $\theta_i$ between two TTIs. In particular, $\beta \rightarrow 0$ implies that $\theta_i$ changes slowly since the probability of a large difference between $\theta_{i, t-1}$ and $\theta_{i, t}$ is negligible when $\beta \rightarrow 0$; on the contrary, $\beta \rightarrow 1$ corresponds to the cases where $\theta_i$ can change significantly between two consecutive TTIs. Additionally, the parameter $\sigma$ is a boundary, and $C_0$ is the normalization coefficient. Furthermore, as in \cite{POMDP} and \cite{My-TWC}, we assume that the complex path gain coefficient $\alpha_i \sim {\cal CN}(0, 1)$ and it changes independently from one TTI to another. Until now, we have introduced the dynamic modeling of the time-varying AoDs, and in the following we elaborate on the employed beam training protocol.

\begin{figure*}[!ht]
\normalsize
\setcounter{MYtempeqncnt}{\value{equation}}
\setcounter{equation}{9}
\begin{eqnarray}\label{SP-2}
    \Pr \left\{[\mathbf e_t]_{\kappa} = 1 \mid \mathbf y_t \right\} & = & \int\limits_{-\infty}^{\infty} \int\limits_{-\infty}^{\infty} \frac{\Pr\left\{[\mathbf e_t]_{\kappa} = 1, \Re\{\alpha_t\}, \Im\{\alpha_t\}, \mathbf y_t \right\}}{\Pr\left\{\mathbf y_t \right\}} ~\text{d} \Re\{\alpha_t\} ~\text{d} \Im\{\alpha_t\} \nonumber \\[2ex]
    & \overset{(a)}{\propto} & \int\limits_{-\infty}^{\infty} \int\limits_{-\infty}^{\infty} \Pr\left\{[\mathbf e_t]_{\kappa} = 1, \Re\{\alpha_t\}, \Im\{\alpha_t\} \right\} \Pr\left\{\mathbf y_t \mid \Re\{\alpha_t\}, \Im\{\alpha_t\}, [\mathbf e_t]_{\kappa} = 1 \right\} ~\text{d} \Re\{\alpha_t\} ~\text{d} \Im\{\alpha_t\} \nonumber \\[2ex]
    & \propto & \int\limits_{-\infty}^{\infty} \int\limits_{-\infty}^{\infty} \underbrace{\Pr\left\{[\mathbf e_t]_{\kappa} = 1 \right\}}_{[\bm \eta_t^{pri}]_{\kappa}} \exp \left(-|\alpha_t|^2\right) \exp\left(-\gamma \|\mathbf y_t - \alpha_t [{\bf S}_t]_{:, \kappa}\|^2 \right) ~\text{d} \Re\{\alpha_t\} ~\text{d} \Im\{\alpha_t\} \nonumber \\[2ex]
    & \propto & \frac{[\bm \eta_t^{pri}]_{\kappa} \exp\left(-{\mathbf y_t}^{\rm H} {\bm \Sigma^{-1}_{\kappa,t}} {\mathbf y_t}\right)}{|\bm \Sigma_{\kappa,t}|} ~\triangleq~ [\bm \eta_t^{post}]_{\kappa}.
\end{eqnarray}
\hrulefill
\vskip -1em
\end{figure*}
\setcounter{equation}{\value{MYtempeqncnt}}

As shown in Fig. \ref{BT-Protocol}, we focus on one transmission frame which consists of $P$ TTIs. In the initial TTI, since no useful information about the AoDs can be leveraged, we assign $M_{ini}$ channel uses for conventional beam training. While for the subsequent $(P-1)$ TTIs, since the AoDs estimated in the previous TTI and the AoD transition function can be leveraged, we only assign $M$ channel uses for beam tracking, where $M < M_{ini}$. If the $M$ well-designed training beams can achieve a comparable performance against the conventional $M_{ini}$ ones, more channel uses are therefore left for data communication, leading to a higher throughput. Last, since various conventional beam training algorithms have been investigated in the open literature, e.g., \cite{MAP}, we thus only consider the beam tracking periods in this paper.

\section{Training Beam Sequence Design}\label{TBD}
From this section, we focus on the single-path channel model, i.e., $L = 1$, and introduce the proposed training beam sequence design approach. Note that the path index $i$ is dropped for brevity.

\subsection{Formulated Sparse Recovery Problem}
Considering the $t$-th beam training period, $\forall t \in \{2, \cdots, P\}$, if we denote the $m$-th pilot symbol by $x_{m,t} = 1$, and the corresponding transmit beamforming vector by ${\bf f}_{m,t}$, the received signal is then given by
\begin{equation}\label{CS-1}
    y_{m,t} = {\bf h}_t^{\rm H} {\bf f}_{m,t} + n_{m,t},
\end{equation}
where $n_{m,t} \sim {\cal CN}(0, \frac{1}{\gamma})$ is the additive Gaussian noise and $\gamma$ is the training signal-to-noise ratio (SNR). We then introduce a beam codebook matrix
\begin{equation}
    {\bf A}_{\rm T} = \left[\mathbf a_{\rm T}\left(\bar \theta_1\right), \cdots, \mathbf a_{\rm T}\left(\bar \theta_N\right)\right],
\end{equation}
and transform ${\bf h}_t$ into its beamspace representation, given by
\begin{equation}\label{CS-2}
    {\bf h}_t = \sqrt{N_{\rm T}} \alpha_t {\bf a}_{\rm T}(\theta_t) = \sqrt{N_{\rm T}} \alpha_t {\bf A}_{\rm T} {\bf e}_t,
\end{equation}
where the vector $\mathbf e_t$ has only one nonzero element whose location is uniquely determined by $\theta_t$ according to our previous on-grid assumption. By substituting \eqref{CS-2} into \eqref{CS-1}, we rewrite the latter as
\begin{equation}\label{CS-3}
    y_{m,t}^\ast = \sqrt{N_{\rm T}} \alpha_t \mathbf f_{m,t}^{\rm H} {\bf A}_{\rm T} {\bf e}_t + n_{m,t}^\ast.
\end{equation}
Stacking the $M$ received pilot symbols into a vector ${\bf y}_t = [y_{1,t}, \cdots, y_{M,t}]^{\rm H}$, we then construct a sparse recovery problem, given by
\begin{equation}\label{CS-4}
    {\bf y}_t = \alpha_t {\bf S}_t {\bf e}_t + {\bf n}_t,
\end{equation}
where ${\bf S}_t = \sqrt{N_{\rm T}}{\bf F}_t^{\rm H} {\bf A}_{\rm T}$ is termed sensing matrix with ${\bf F}_t = [{\bf f}_{1,t}, \cdots, {\bf f}_{M,t}]$, and ${\bf n}_t = [n_{1,t}, \cdots, n_{M,t}]^{\rm H}$.

\subsection{Proposed MAP Estimation}
In this section, we propose to use the MAP criterion to estimate the position of the nonzero entry in $\mathbf e_t$. Without loss of generality, we assume that the current AoD is $\theta_t = \bar \theta_{\kappa}$, such that $[\mathbf e_t]_{\kappa} = 1$ according to \eqref{CS-2}. Thus, we can recast \eqref{CS-4} as
\begin{equation}\label{SP-1}
    {\bf y}_t = \alpha_t [{\bf S}_t]_{:, \kappa} + {\bf n}_t.
\end{equation}
We then compute the posterior probability of $[{\bf e}_t]_{\kappa} = 1$, given by \eqref{SP-2}, where $\Pr(\mathbf y_t)$ is omitted in $(a)$ since it is a constant in each entry of $\bm \eta^{post}_t$, and $\bm \Sigma_{\kappa,t} = [{\bf S}_t]_{:, \kappa} [{\bf S}_t]^{\rm H}_{:, \kappa} + \frac{1}{\gamma}{\bf I}$. To facilitate subsequent computations, we normalize the summation of $\bm \eta^{post}_t$ to one. Moreover, to incorporate the previous tracking errors, we compute the prior probability of $[\mathbf e_t]_{\kappa} = 1$ in a recursive fashion, given by
\setcounter{equation}{10}
\begin{equation}\label{SP-3}
    [\bm \eta_t^{pri}]_{\kappa} = \sum\limits_{\iota = 1}^{N} \Pr\left\{[\mathbf e_t]_{\kappa} = 1 ~|~ [\mathbf e_{t-1}]_{\iota} = 1\right\} [\bm \eta^{post}_{t-1}]_{\iota},
\end{equation}
where $\Pr\left\{[\mathbf e_t]_{\kappa} = 1 \mid [\mathbf e_{t-1}]_{\iota} = 1\right\}$ is the transition probability from $[\mathbf e_{t-1}]_{\iota} = 1$ to $[\mathbf e_t]_{\kappa} = 1$, or equivalently, from $\theta_{t-1} = \bar \theta_{\iota}$ to $\theta_{t} = \bar \theta_{\kappa}$. According to the MAP criterion, the estimated location of the nonzero element in ${\bf e}_t$ is given by
\begin{equation}\label{SP-4}
  \hat \kappa = \argmax_{n = 1, \cdots, N} ~\big\{\bm \eta^{post}_t\big\}.
\end{equation}

According to \cite{PF-TSP}, \eqref{SP-4} provides the optimal estimation of AoD in each beam tracking period. Last, once AoD estimation is completed, the user feeds $\hat \kappa$ back to the BS for subsequent data communication and the next beam tracking period.

\begin{figure*}[!ht]
\normalsize
\setcounter{MYtempeqncnt}{\value{equation}}
\setcounter{equation}{13}
\begin{eqnarray}\label{UB-4}
    \hspace{-1.7cm} \Gamma_t^{\rm err} & \triangleq & 1 - \sum\limits_{\kappa = 1}^{N} [\bm \eta^{pri}_t]_{\kappa} \Pr\left\{\bigcap\limits_{n = 1, n \ne \kappa}^{N} \Big([\bm \eta^{post}_t]_{\kappa} > [\bm \eta^{post}_t]_{n}\Big) ~\bigg|~ [{\bf e}_t]_{\kappa} = 1 \right\} \nonumber \\[2ex]
    \hspace{-1.7cm} & = & 1 - \sum\limits_{\kappa = 1}^{N} [\bm \eta^{pri}_t]_{\kappa} \left(1 - \Pr\left\{\bigcup\limits_{n = 1, n \ne \kappa}^{N} \Big([\bm \eta^{post}_t]_{\kappa} \le [\bm \eta^{post}_t]_{n}\Big) ~\bigg|~ [{\bf e}_t]_{\kappa} = 1 \right\}\right) \nonumber \\[2ex]
    \hspace{-1.7cm} & \overset{(b)}{=} & \sum\limits_{\kappa = 1}^{N} [\bm \eta^{pri}_t]_{\kappa} \Pr\left\{\bigcup\limits_{n = 1, n \ne \kappa}^{N} \Big([\bm \eta^{post}_t]_{\kappa} \le [\bm \eta^{post}_t]_{n}\Big) ~\bigg|~ [{\bf e}_t]_{\kappa} = 1 \right\}.
\end{eqnarray}
\hrulefill
\begin{eqnarray}\label{EER-1}
    \Gamma_t^{\rm err} & \overset{(c)}{\le} & \sum\limits_{\kappa = 1}^{N} [\bm \eta^{pri}_t]_{\kappa} \sum\limits_{n = 1, n \ne \kappa}^{N} \Pr\left\{\Big([\bm \eta^{post}_t]_{\kappa} \le [\bm \eta^{post}_t]_{n}\Big) ~\bigg|~ [{\bf e}_t]_{\kappa} = 1 \right\} \nonumber \\[2ex]
    & = & \underbrace{\sum\limits_{\kappa = 1}^{N} [\bm \eta^{pri}_t]_{\kappa} \sum\limits_{n = 1, n \ne \kappa}^{N} ~\underbrace{\Pr\left\{{\mathbf y_t}^{\rm H} (\bm \Sigma_{n,t}^{-1} - \bm \Sigma_{\kappa,t}^{-1}) {\mathbf y_t} \le \ln \left(\frac{[\bm \eta^{pri}_t]_n}{[\bm \eta^{pri}_t]_{\kappa}} \frac{|\bm \Sigma_{\kappa,t}|}{|\bm \Sigma_{n,t}|} \right) ~\bigg|~ [{\bf e}_t]_{\kappa} = 1 \right\}}_{\mu_{\kappa, n, t}}}_{\Gamma_t^{\rm err, ub}}.
\end{eqnarray}
\hrulefill
\vskip -1em
\end{figure*}
\setcounter{equation}{\value{MYtempeqncnt}}

\subsection{Beam Tracking Error Probability}\label{UB}
Before introducing the proposed training beam sequence design approach, we first discuss the metric used for evaluating the performance of a particular set of training beams. In accordance with our assumption in \eqref{SP-1} that the $\kappa$-th entry in ${\bf e}_t$ is nonzero, under the proposed MAP criterion in \eqref{SP-4}, the probability of successfully estimating the current AoD can be expressed as
\begin{equation*}\label{UB-1}
    \Gamma_{\kappa, t}^{\rm se} =  \Pr\left\{\bigcap\limits_{n = 1, n \ne \kappa}^{N} \Big([\bm \eta^{post}_t]_{\kappa} > [\bm \eta^{post}_t]_n\Big) ~\bigg|~ [{\bf e}_t]_{\kappa} = 1 \right\}.
\end{equation*}
Recall that $\kappa \in \{1, \cdots, N\}$. Taking each possible case of $\kappa$ into consideration, the expected probability of successfully estimating $\theta_t$ is then given by
\begin{eqnarray}\label{UB-2}
    \Gamma_t^{\rm se} & = & \sum\limits_{\kappa = 1}^{N} [\bm \eta^{pri}_t]_{\kappa} ~ \Gamma_{\kappa, t}^{\rm se}.
\end{eqnarray}
The TEP in the $t$-th TTI is then given by \eqref{UB-4}, where $(b)$ is due to $\sum\nolimits_{\kappa = 1}^{N} [\bm \eta^{pri}_t]_{\kappa} = 1$.

\begin{figure*}[!ht]
\normalsize
\setcounter{MYtempeqncnt}{\value{equation}}
\setcounter{equation}{17}
\begin{eqnarray}\label{Mu-1}
    \mu_{\kappa, n, t} & = & \Pr\Big\{{\bf y}_t^{\rm H} (\bm \Sigma_{n, t}^{-1} - \bm \Sigma_{\kappa, t}^{-1}) {\bf y}_t \le \delta_{\kappa, n, t} ~\big|~ [{\bf e}_t]_{\kappa} = 1 \Big\} \nonumber \\[3ex]
    & = & \Pr\Big\{\lambda_{\kappa, n, t, 1} \left|[{\bf \ddot{y}}_t]_1\right|^2 + \lambda_{\kappa, n, t, 2} \left|[{\bf \ddot{y}}_t]_2\right|^2 \le \delta_{\kappa, n, t} ~\big|~ [{\bf e}_t]_{\kappa} = 1 \Big\} \nonumber \\[3ex]
    & = & \begin{dcases}
    \frac{\lambda_{\kappa, n, t, 2}}{\lambda_{\kappa, n, t, 2} - \lambda_{\kappa, n, t, 1}} \exp\left(-\frac{\delta_{\kappa, n, t}}{\lambda_{\kappa, n, t, 2}}\right), & \delta_{\kappa, n, t} \le 0, \\[2ex]
    1 - \frac{\lambda_{\kappa, n, t, 1}}{\lambda_{\kappa, n, t, 1} - \lambda_{\kappa, n, t, 2}} \exp\left(-\frac{\delta_{\kappa, n, t}}{\lambda_{\kappa, n, t, 1}}\right), & \delta_{\kappa, n, t} > 0.
    \end{dcases}
\end{eqnarray}
\hrulefill
\vskip -1em
\end{figure*}
\setcounter{equation}{\value{MYtempeqncnt}}

\subsection{A Closed-Form Upper Bound for TEP}\label{ATEP-UB}
Expressing \eqref{UB-4} in an explicit form is challenging, and hence we turn to its upper bound, given by \eqref{EER-1}, where $(c)$ is due to the Boole's inequality and $\bm \Sigma_{n,t} = [{\bf S}_t]_{:, n} [{\bf S}_t]^{\rm H}_{:, n} + \frac{1}{\gamma}{\bf I}$. To derive an explicit expression for $\displaystyle \Gamma_t^{\rm err, ub}$, we need to first determine the distribution of ${\mathbf y_t}^{\rm H} (\bm \Sigma_{n,t}^{-1} - \bm \Sigma_{\kappa,t}^{-1}) {\mathbf y_t}$.

Recall that ${\mathbf y_t} \sim {\cal CN}({\mathbf 0}, \bm \Sigma_{\kappa,t})$ when $[\mathbf e_t]_{\kappa} = 1$. By leveraging the eigenvalue decomposition of $\bm \Sigma_{\kappa,t}$, given by
\begin{equation*}
    \bm \Sigma_{\kappa,t} = [{\bf S}_t]_{:, \kappa} [{\bf S}_t]_{:, \kappa}^{\rm H} + \frac{1}{\gamma} {\bf I} = {\bf U}_{\kappa,t} \bm \Delta_{\kappa,t} {\bf U}_{\kappa,t}^{\rm H},
\end{equation*}
we can rewrite $\mathbf y_t$ as
\begin{equation*}
    {\mathbf y_t} = {\bf U}_{\kappa,t} {\bm \Delta}_{\kappa,t}^{\frac{1}{2}} {\bf \dot{y}}_t,
\end{equation*}
and transform ${\mathbf y_t}^{\rm H} (\bm \Sigma_{n,t}^{-1} - \bm \Sigma_{\kappa,t}^{-1}) {\mathbf y_t}$ into
\begin{eqnarray*}
{\mathbf y}_t^{\rm H} (\bm \Sigma_{n,t}^{-1} - \bm \Sigma_{\kappa,t}^{-1}) {\mathbf y_t} ~~~~~~~~~~~~~~~~~~~~~~~~~~~~~~~~~~~ \\[2ex]
= {\bf \dot{y}}_t^{\rm H} {\bm \Delta}_{\kappa,t}^{\frac{1}{2}} {\bf U}_{\kappa,t}^{\rm H} (\bm \Sigma_{n,t}^{-1} - \bm \Sigma_{\kappa,t}^{-1}) {\bf U}_{\kappa,t} {\bm \Delta}_{\kappa,t}^{\frac{1}{2}} {\bf \dot{y}}_t,
\end{eqnarray*}
where ${\bf \dot{y}}_t \sim {\cal CN}(\mathbf 0, \mathbf I)$. Denoting the eigenvalue decomposition of ${\bm \Delta}_{\kappa,t}^{\frac{1}{2}} {\bf U}_{\kappa,t}^{\rm H} (\bm \Sigma_{n,t}^{-1} - \bm \Sigma_{\kappa,t}^{-1}) {\bf U}_{\kappa,t} {\bm \Delta}_{\kappa,t}^{\frac{1}{2}}$ by ${\bf V}_t {\bm \Lambda}_{\kappa, n, t} {\bf V}_t^{\rm H}$, we can further transform ${\bf \dot{y}}_t^{\rm H} {\bm \Delta}_{\kappa,t}^{\frac{1}{2}} {\bf U}_{\kappa,t}^{\rm H} (\bm \Sigma_{n,t}^{-1} - \bm \Sigma_{\kappa,t}^{-1}) {\bf U}_{\kappa,t} {\bm \Delta}_{\kappa,t}^{\frac{1}{2}} {\bf \dot{y}}_t$ into
\begin{eqnarray*}
    {\bf \dot{y}}_t^{\rm H} {\bm \Delta}_{\kappa,t}^{\frac{1}{2}} {\bf U}_{\kappa,t}^{\rm H} (\bm \Sigma_{n,t}^{-1} - \bm \Sigma_{\kappa,t}^{-1}) {\bf U}_{\kappa,t} {\bm \Delta}_{\kappa,t}^{\frac{1}{2}} {\bf \dot{y}}_t ~~~~~~~~~~ \\[2ex]
    ~=~ {\bf \dot{y}}_t^{\rm H} {\bf V}_t {\bm \Lambda}_{\kappa, n, t} {\bf V}_t^{\rm H} {\bf \dot{y}}_t ~=~ {\bf \ddot{y}}_t^{\rm H} {\bm \Lambda}_{\kappa, n, t} {\bf \ddot{y}}_t,
\end{eqnarray*}
where ${\bf \ddot{y}}_t = {\bf V}_t^{\rm H} {\bf \dot{y}}_t$ and it follows ${\cal CN}(\bm 0, \mathbf I)$ since $\mathbf V_t$ is an unitary matrix. Until now, we have successfully transformed ${\bf y}_t^{\rm H} (\bm \Sigma_{n,t}^{-1} - \bm \Sigma_{\kappa,t}^{-1}) {\bf y}_t$ into
\setcounter{equation}{15}
\begin{eqnarray}\label{EER-2}
    {\bf y}_t^{\rm H} (\bm \Sigma_{n,t}^{-1} - \bm \Sigma_{\kappa,t}^{-1}) {\bf y}_t & = & {\bf \ddot{y}}_t^{\rm H} {\bm \Lambda}_{\kappa, n, t} {\bf \ddot{y}}_t \nonumber \\[2ex]
    & = & \sum\limits_{i = 1}^M \lambda_{\kappa, n, t, i} \left|[{\bf \ddot{y}}_t]_i\right|^2,
\end{eqnarray}
where $\lambda_{\kappa, n, t, i}$ is the $i$-th largest diagonal entry in $\bm \Lambda_{\kappa, n, t}$. Furthermore, by rewriting $\bm \Sigma_{n, t}^{-1} - \bm \Sigma_{\kappa, t}^{-1}$ as
\begin{eqnarray}
    \bm \Sigma_{n,t}^{-1} - \bm \Sigma_{\kappa,t}^{-1} = \frac{[{\bf S}_t]_{:, \kappa} [{\bf S}_t]_{:, \kappa}^{\rm H}}{\frac{1}{\gamma^2} + \frac{1}{\gamma}\|[{\bf S}]_{:, \kappa}\|^2} - \frac{[{\bf S}_t]_{:, n} [{\bf S}_t]_{:, n}^{\rm H}}{\frac{1}{\gamma^2} + \frac{1}{\gamma}\|[{\bf S}_t]_{:, n}\|^2},
\end{eqnarray}
we can realize that $rank(\bm \Sigma_{n,t}^{-1} - \bm \Sigma_{\kappa,t}^{-1}) \le 2$, and consequently $rank({\bm \Lambda}_{\kappa, n, t}) \le 2$, which implies that $\bm \Lambda_{\kappa, n, t}$ has at most two nonzero diagonal entries. Moreover, since ${\bf \ddot{y}}_t \sim {\cal CN}(\mathbf 0, \mathbf I)$, both $\left|[{\bf \ddot{y}}_t]_1\right|^2$ and $\left|[{\bf \ddot{y}}_t]_2\right|^2$ follow an exponential distribution with mean $1$. Thanks to this observation, we can derive a closed-form expression for $\mu_{\kappa, n, t}$, as detailed below.

\hspace{-0.35cm}\textbf{Case 1}: If $\lambda_{\kappa, n, t, 1} > 0$, $\lambda_{\kappa, n, t, 2} < 0$, we can express $\mu_{\kappa, n, t}$ as \eqref{Mu-1}, where $\delta_{\kappa, n, t} = \ln \left(\frac{[\bm \eta^{pri}_t]_n}{[\bm \eta^{pri}_t]_{\kappa}} \frac{|\bm \Sigma_{\kappa,t}|}{|\bm \Sigma_{n,t}|} \right)$.

\hspace{-0.35cm}\textbf{Case 2}: $\lambda_{\kappa, n, t, 1} > 0$ and $\lambda_{\kappa, n, t, 2} = 0$,
\begin{equation*}
    \mu_{\kappa, n, t} ~=~ \begin{dcases}
    1 - \exp\left(-\frac{\delta_{\kappa, n, t}}{\lambda_{\kappa, n, t, 1}}\right), & \delta_{\kappa, n, t} > 0, \\[2ex]
    0, & \delta_{\kappa, n, t} \le 0.
    \end{dcases}
\end{equation*}

\hspace{-0.35cm}\textbf{Case 3}: $\lambda_{\kappa, n, t, 1} = 0$ and $\lambda_{\kappa, n, t, 2} < 0$,
\begin{equation*}
    \mu_{\kappa, n, t} ~=~ \begin{dcases}
    \exp\left(-\frac{\delta_{\kappa, n, t}}{\lambda_{\kappa, n, t, 2}}\right), ~~~~~ & \delta_{\kappa, n, t} < 0, \\[2ex]
    1, ~~~~~ & \delta_{\kappa, n, t} \ge 0.
    \end{dcases}
\end{equation*}

\hspace{-0.35cm}\textbf{Case 4}: $\lambda_{\kappa, n, t, 1} = 0$ and $\lambda_{\kappa, n, t, 2} = 0$,
\begin{equation*}
    \hspace{-2.95cm} \mu_{\kappa, n, t} ~=~ \begin{dcases}
    0, & \delta_{\kappa, n, t} < 0, \\[3ex]
    1, & \delta_{\kappa, n, t} > 0.
    \end{dcases}
\end{equation*}

\subsection{Formulated Optimization Problem}
As shown in \eqref{SP-2} and \eqref{SP-3}, the $M$ training beams used in the $(t-1)$-th beam tracking period also influence the $t$-th beam tracking period via $\bm \eta_t^{pri}$. Nonetheless, we can still focus on one beam tracking period and optimize its associated $M$ training beams to minimize the TEP, since the more accurate the current beam tracking result is, the better the subsequent beam tracking result can be.

Therefore, we employ $\displaystyle \Gamma_t^{\rm err, ub}$ as the objective function and formulate the training beam sequence design problem as
\setcounter{equation}{18}
\begin{eqnarray}
    \textbf{P}_1: & \min\limits_{\mathbf F_t} & \Gamma_t^{\rm err, ub} \label{OP-0A} \\
    & \text{s.t.} & |[\mathbf F_t]_{n,m}| = \frac{1}{\sqrt{N_{\rm T}}}, \label{OP-0B} \\[2ex]
    & & \forall n \in \{1, \cdots, N_{\rm T}\}, ~m \in \{1, \cdots, M\}, \nonumber
\end{eqnarray}
where \eqref{OP-0B} accounts for the unit modulus constraint imposed by analog phase shifters \cite{RH-Survey}. Due to the complicated objective function $\displaystyle \Gamma_t^{\rm err, ub}$ and the highly non-convex constraint \eqref{OP-0B}, we resort to PSA\footnote{PSA is an evolutionary computation technique and it searches for optima by updating generations rather than by gradient. As such, PSA can be used to solve very difficult problems such as the ones with non-differentiable objective functions. Despite the powerfulness of PSA, it cannot guarantee an optimal solution is found. For more details about PSA, refer to \cite{PSO}.} to solve $\textbf{P}_1$.

\textbf{Remark}: The proposed beam tracking approach for the single-path channel can be readily extended to multi-path scenarios. Specifically, we can track the multiple paths in an iterative manner, and only one path is addressed within each iteration. Following the idea of successive interference cancellation, the contributions of the paths that have been found in the previous iterations should be subtracted from the received signals before finding new path.

\section{Numerical Results}\label{NR}
In this section, we provide numerical results to evaluate the performance of the proposed beam tracking approach. In simulations, the BS is equipped with $N_{\rm T} = 32$ antennas and all these antennas are connected to a single radio frequency chain, leading to an analog-only architecture. Following \cite{Duan-Tracking} and \cite{POMDP}, we assume that one transmission frame consists of $P = 10$ TTIs and the AoD in the first TTI is accurately estimated via conventional beam training. The other parameters are defined as follows: $N = 64$, $M = 2$, and $\sigma = 5$.

\begin{figure}
\vskip -10pt
\centering
\includegraphics[width = 8.2cm]{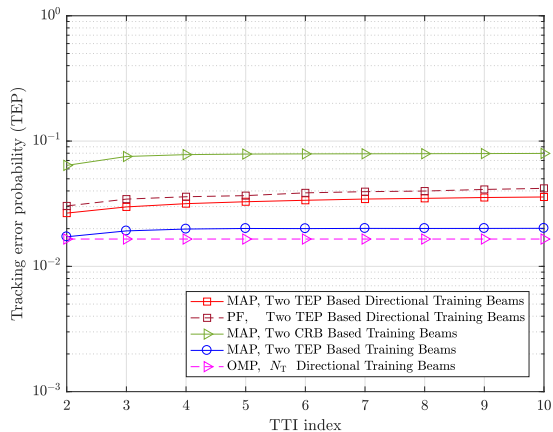}
\caption{TEP with respect to TTI index, where $\beta = 0.2$ and $\gamma = 10$dB.}\label{SP-TEP-Time}
\end{figure}

The performance of our proposed beam tracking approach is shown in Fig. \ref{SP-TEP-Time}, where four benchmark approaches are also provided for comparison: 1) MAP with two TEP based directional training beams, i.e., the two beam codewords in $\mathbf A_{\rm T}$ that minimize $\displaystyle \Gamma_t^{\rm err, ub}$; 2) Particle filtering (PF) estimator with two TEP based directional training beams; 3) MAP with two CRB based training beams \cite{CRLB}; and 4) Orthogonal matching pursuit (OMP) with $N_{\rm T}$ directional training beams \cite{MAP}. From this figure, we easily observe that the proposed TEP based training beams perform better than the CRB based training beams and the TEP based directional training beams. In particular, when $\beta = 0.2$, by leveraging MAP estimation and the proposed TEP based training beam sequence design, we only need to assign two channel uses for beam tracking. The resultant TEP performance is comparable to that of the beam cycling approach which consumes $N_{\rm T}$ channel uses for conventional beam training at the beginning of each TTI.

\begin{figure}
\vskip -10pt
\centering
\includegraphics[width = 8.2cm]{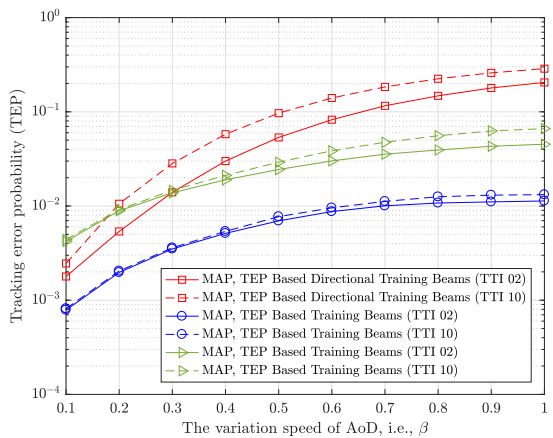}
\caption{TEP with respect to different values of $\beta$, where $\gamma = 20$dB.}\label{SP-TEP-Beta}
\end{figure}

The TEP performances of our proposed beam tracking approach in different time-varying scenarios are shown in Fig. \ref{SP-TEP-Beta}, where the performances of the CRB based training beams and the TEP based directional training beams are also provided. From this figure, we immediately observe that our proposed TEP based training beam sequence design approach performs better than the other two benchmark schemes across the whole range of $\beta$, which further validates the superiority of our proposed TEP based training beam sequence design approach. We can also observe from this figure that the performance gap between the TEP based training beams and the TEP based directional training beams becomes larger as $\beta$ increases. The reason is that directional training beams have narrow beamwidths and hence the maximum number of AoD candidate directions they can probe is limited compared to the training beams generated by $\textbf{P}_1$. When $\beta$ is large, which means that the AoD can change significantly between two consecutive TTIs, since the $M = 2$ directional training beams can cover only a few AoD candidate directions, the current AoD will be missed if it is located in the uncovered directions. In contrast, the training beams generated by $\textbf{P}_1$ are much more flexible. When the AoD changes significantly, the training beams output by $\textbf{P}_1$ are expected to have wide beamwidth in order to cover more AoD candidate directions, thereby significantly reducing the beam tracking error.

\begin{figure}
\vskip -10pt
\centering
\includegraphics[width = 8.2cm]{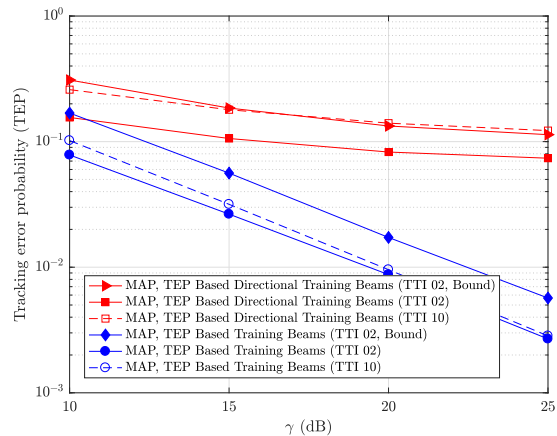}
\caption{TEP with respect to the training SNR, where $\beta = 0.6$.}\label{SP-TEP-SNR}
\end{figure}

The performances of our proposed beam tracking approach under different SNR scenarios are provided in Fig. \ref{SP-TEP-SNR}, where the performances of the TEP based directional training beams are also provided for comparison. From this figure, it is observed that our derived TEP upper bound $\displaystyle \Gamma_t^{\rm err, ub}$ decreases following a similar trend as the simulated TEP in the second TTI when increasing the training SNR from 10dB to 25dB, which validates the effectiveness of $\displaystyle \Gamma_t^{\rm err, ub}$ as the objective function to optimize the training beam sequence. We can also observe from this figure that our proposed TEP based training beams perform much better than the TEP based directional training beams and their performance gap becomes larger as SNR increases. Moreover, as shown in Fig. \ref{SP-TEP-SNR}, the performance of the TEP based directional training beams has an ``error floor''. As explained earlier, this is due to the narrow beamwidth of directional training beams. Last, for the proposed beam tracking approach, we also observe that its performance gap between the second and the last TTIs becomes narrower as the training SNR increases from 10dB to 25dB.

\section{Conclusions}\label{CN}
In this paper, we considered a mmWave MISO channel and imposed a discrete Markov process assumption on the evolution of its time-varying AoD. To incorporate the AoD transition information into beam tracking design, we introduced the MAP criterion for AoD update in each beam tracking period. We derived a closed-form upper bound for the TEP, and leveraged it as the objective function to optimize the training beam sequence used in each beam tracking period. Numerical results validated the good performance of the proposed beam tracking approach for sparse mmWave channels.

\end{document}